\documentclass[aps,twocolumn,showpacs,preprintnumbers]{revtex4}

\usepackage{graphicx}
\usepackage{dcolumn}
\usepackage{bm}
\usepackage{hyperref}

\newcommand{\beq}{\begin{equation}}
\newcommand{\eeq}{\end{equation}}
\newcommand{\beqa}{\begin{eqnarray}}
\newcommand{\eeqa}{\end{eqnarray}}
\newcommand{\beqar}{\begin{eqnarray*}}
\newcommand{\eeqar}{\end{eqnarray*}}
\newcommand{\bra}[1]{\mbox{$\langle{#1}|$}}
\newcommand{\ket}[1]{\mbox{$|{#1}\rangle$}}

\def\Tr{{\rm Tr}}

\newcounter{saveeqn}
\newcommand{\alpheqn}{\setcounter{saveeqn}{\value{equation}}%
\stepcounter{saveeqn}\setcounter{equation}{0}%
\renewcommand{\theequation}{\mbox{\arabic{saveeqn}\alph{equation}}}}
\newcommand{\reseteqn}{\setcounter{equation}{\value{saveeqn}}%
\renewcommand{\theequation}{\arabic{equation}}}
\def\beql{\alpheqn \beqa}
\def\eeql{\eeqa \reseteqn}

\begin{document}

\title{Separability criterion  for pure states in multipartite and high dimensional systems}
\author{An Min WANG$^{1,2}$}

\altaffiliation{Supported by the National Natural Science Foundation of China under Grant No. 60173047, ``973" Project of China and the Natural Science Foundation of Anhui Province}

\affiliation{$^{1}$ Laboratory of Quantum Communication and Quantum Computing
and Institute for Theoretical Physics}
\affiliation{$^{2}$Department of Modern Physics,
University of Science and Technology of China \\
P.O. Box 4, Hefei 230027, People's Republic of China}



\begin{abstract}
We propose a sufficient and necessary separability criterion for pure states in multipartite and high dimensional systems. Its main advantage is operational and computable. The obvious expressions of this criterion can be given out by the coefficients of components of the pure state. In the end, we simply mention a principle method how to define and obtain the measures of entanglement in multipartite and high dimensional systems. 
\end{abstract}


\maketitle

Entanglement is one of the most striking feature of quantum theory.  It is very important how to quantify it. In our point of view,  separability is a theoretical foot stone to define the measures of entanglement. 

A $n$-partite quantum state $\rho_{A_1\!A_2\!\cdots\! A_n}$ is called (fully) separable iff it can be written as a convex combination of product states, {\it i.e.}
\beq
\label{sdef}
\rho_{A_1\!A_2\!\cdots\! A_n}=\sum_m p_m \rho_{A_1}^{(m)}\otimes\rho_{A_2}^{(m)}\otimes\cdots\otimes\rho_{A_n}^{(m)}
\eeq
where any $\rho_{A_i}^{(m)}\  (i=1,2,\cdots,n)$ is a density matrix of $r_i$-dimensional. Otherwise, $\rho_{A_1\!A_2\!\cdots\! A_n}$ is entangled. For low dimensional ($2\times 2$ and $2\times 3$) of bipartite systems, Peres and Horodecki gave out a necessary and sufficient condition for separability of arbitrary states, that is, $\rho_{A_1\!A_2}$ is separable if and only if it has a ``positive partial transpose" (PPT) \cite{Peres,Horodecki1}. Unfortunately, this criterion is only necessary for a sate to be separable in higher dimensions, but is not sufficient. Recently a lot of works about separability have been presented \cite{Lewenstein,Rudolph,Chen,Horodecki2,Doherty,Giedke,Karnas,Dur,Acin,My1}.  These works explain a fact that the operational and computable separability criterion, in particular, in multipartite and high dimensional systems becomes very interesting now.  Here, our letter will devote to this problem. 

First of all, we need the following lemma:

{\bf Lemma}\ If a pure state $\rho_{A_1\!A_2\!\cdots\! A_n}^{\rm P}$ is separable, thus it must be ``purely" separable, that is
\beq
\rho_{A_1\!A_2\!\cdots\! A_n}^{\rm P}=\ket{\psi_{A_1\!A_2\!\cdots\! A_n}}\bra{\psi_{A_1\!A_2\!\cdots\! A_n}}=\prod_{i=1,\otimes}^n\rho_{A_i}
\eeq

{\bf Proof}\ Sine $\rho_{A_1\!A_2\!\cdots\! A_n}^{\rm P}$ is separable, it must be able to write as the form of eq.(\ref{sdef}). Obviously, for a given $m$, $\rho_{A_1}^{(m)}\otimes\rho_{A_2}^{(m)}\otimes\cdots\otimes\rho_{A_n}^{(m)}$, as  a Hermit operator in Hilbert space ${\cal{H}}_{A_1}\otimes{\cal{H}}_{A_2}\otimes\cdots\otimes{\cal{H}}_{A_n}$, has a  complete set of orthogonal and normalized eigenvectors, which is denoted by $\{\ket{\psi_{A_1\!A_2\!\cdots\! A_n}^m(\alpha)}, (\alpha=1,2,\cdots,N)\}$.  When a value of $m$ is chosen, we can take them as a set of basis of Hilbert space ${\cal{H}}_{A_1}\otimes{\cal{H}}_{A_2}\otimes\cdots\otimes{\cal{H}}_{A_n}$, it implies we can write
\beq
\ket{\psi_{A_1\!A_2\!\cdots\! A_n}}=\sum_{\alpha=1}^N C_\alpha^m \ket{\psi_{A_1\!A_2\!\cdots\! A_n}^m(\alpha)}
\eeq
Note that a $m$-th component state $\rho_{A_1}^{(m)}\otimes\rho_{A_2}^{(m)}\otimes\cdots\otimes\rho_{A_n}^{(m)}$ has been taken as a pure state, without loss of generality, we set 
$ \prod_{i=1,\otimes}^n\rho_{A_i}^{(m)}=\ket{\psi_{A_1\!A_2\!\cdots\! A_n}^m(1)}\bra{\psi_{A_1\!A_2\!\cdots\! A_n}^m(1)}$. 
Thus, 
\beqa
\!\!1&\!=\!&\bra{\psi_{A_1\!A_2\!\cdots\! A_n}} \rho_{A_1\!A_2\!\cdots\! A_n}^{\rm P}\ket{\psi_{A_1\!A_2\!\cdots A_n}}\nonumber\\
&\!=\!&\sum_{m}p_m \bra{\psi_{A_1\!A_2\!\cdots\! A_n}} \prod_{i=1,\otimes}^n\rho_{A_i}^{(m)}\ket{\psi_{A_1\!A_2\!\cdots\! A_n}}\nonumber\\
&\! =\!&\!\!\sum_{m,\alpha,\alpha^\prime}\!\!p_mC^{m*}_\alpha C^m_{\alpha^\prime}\bra{\psi^m_{A_1\!A_2\!\cdots\! A_n}(\alpha)} \!\prod_{i=1,\otimes}^n\!\rho_{A_i}^{(m)}\ket{\psi^m_{A_1\!A_2\!\cdots\! A_n}(\alpha^\prime)}\nonumber\\
&\!=\!& \sum_{m} p_m\sum_{\alpha,\alpha^\prime}C^{m*}_\alpha C^m_{\alpha^\prime}\delta_{\alpha\alpha^\prime}\delta_{\alpha^\prime 1}=\sum_{m}p_m C_1^m{}^*C_1^m
\eeqa
It implies that  $C_1^m{}^*C_1^m=1$ and then the other $C^m_\alpha (\alpha=2,3,\cdots,N)$ are zero. Furthermore, we have  $\rho_{A_1}^{(m)}\otimes\rho_{A_2}^{(m)}\otimes\cdots\otimes\rho_{A_n}^{(m)}=\rho_{A_1\!A_2\!\cdots\! A_n}^{\rm P}$, that is, this lemma is valid. 

Without loss of generality, the density matrix  $\rho_{A_1\!A_2\!\cdots\! A_n}$ in a $n$-partite system  can be expanded as
\beqa
\label{eforn}
\rho_{A_1\!A_2\!\cdots\! A_n}&=&\frac{1}{2^n}\sum_{\mu_1=0}^{r_1^2-1}\sum_{\mu_2=0}^{r_2^2-1}\cdots\sum_{\mu_n=0}^{r_n^2-1} \nonumber\\
& &a_{\mu_1\mu_2\cdots \mu_n} \lambda_{A_1}^{\mu_1}\otimes\lambda_{A_2}^{\mu_2}\otimes\cdots\otimes\lambda_{A_n}^{\mu_n}
\eeqa
where $a_{\mu_1\mu_2\cdots \mu_n}$ is a $n$-rank real tensor, $\lambda_{A_i}^0$ is proportional to $r_i$-dimensional  identity matrix and its proportional factor can be taken as $2/r_i$ in usual. While $\lambda_{A_i}^{m_i}\ (m_i=1,2,\cdots, r^2_i-1)$ are generators of $SU(r_i)$. 

For simplicity, we will limit our focus on such a multipartite system in which every partite only has a qubit or/and a qutrit,  and then the corresponding  generators are respectively Pauli matrices $\sigma_i (i=1,2,3)$ and Gell-Mann matrices $\lambda_i\ (i=1,2,3,4,5,6,7,8)$.

{\bf Theorem One.}\  The sufficient and necessary conditions  for separability of an arbitrary pure state in $n$-partite systems made up of $n$ qubits (every partite has a qubit) are 
\beq
\label{scformp}
\bm{\xi}_{A_1}^2=\bm{\xi}_{A_2}^2=\cdots=\bm{\xi}_{A_n}^2=1
\eeq
where $\bm{\xi}_{A_i}$ is a polarized vector of the reduced density matrix $\rho_{A_i}=\Tr_{\prod_{j=1,j\neq i}^n A_j}(\rho_{A_1\!A_2\cdots A_n})=\frac{1}{2}(\sigma_0+\bm{\xi}_{A_i}\cdot\bm{\sigma})$ (tracing off $n-1$ partites).   

{\bf Proof}\ In order to prove it, let us start from the well-known bipartite systems. If denoting the pure state in a bipartite system as
\beq
\label{psi2q}
\ket{\psi_{A_1\!A_2}}=a\ket{00}+b\ket{01}+c\ket{10}+d\ket{11}
\eeq
we can obtain 
\beq
\label{2qpvn2}
\bm{\xi}_{A_1}^2=\bm{\xi}_{A_2}^2=\sum_{i=1}^3 a_{i0} a_{i0}=\sum_{i=1}^3 a_{0i} a_{0i}=1-4|ad-bc|^2
\eeq
Thus, the sufficient and necessary condition  for separability of an arbitrary pure state in bipartite systems of two qubits  can be 
written in form
\begin{equation}
ad=bc  
\label{scfor2q}
\end{equation}

It is easy to prove that theorem one is a necessary creiterion. In fact, if the pure state $\ket{\psi_{A_1\!A_2}}$ is separable, in terms of our lemma, without loss of generality we have 
\begin{equation}
\ket{\psi_{A_1\!A_2}}=(a_1\ket{0}+b_1\ket{1})\otimes(a_2\ket{0}+b_2\ket{1})
\end{equation}
Comparing it with the equation (\ref{psi2q}), up to an undetermined overall phase factor, we can obtain 
\begin{equation}
\label{2qsf}
a=a_1a_2,\quad b=a_1b_2,\quad c=a_2b_1,\quad d=b_1b_2
\end{equation}
Substituting eq.(\ref{2qsf}) into eq.(\ref{2qpvn2}), it immediately following that  our separability criterion is satisfied. 

 Now let us prove it to be sufficient.  If there is only one non-zero element in the coefficient set $\{a,b,c,d\}$ of the concerning pure state (\ref{psi2q})£¬then this state is obviously separable. Actually, at least, there are two non-zero element in the coefficient set, the case is not trivial. It is easy to obtain the product states in all of possible cases as following

(1) $a,b \neq 0$  
\beql
\ket{\psi_{A_1}}&=&\frac{|a|}{\sqrt{|a|^2+|c|^2}}\left(\ket{0}+\frac{c}{a}\ket{1}\right)\\
\ket{\psi_{A_2}}&=&\frac{1}{\sqrt{|a|^2+|b|^2}}(a\ket{0}+b\ket{1})
\eeql

(2) $a,c \neq 0$  
\beql
\ket{\psi_{A_1}}&=&\frac{1}{\sqrt{|a|^2+|c|^2}}\left(a\ket{0}+c\ket{1}\right)\\
\ket{\psi_{A_2}}&=&\frac{|c|}{\sqrt{|c|^2+|d|^2}}\left(\ket{0}+\frac{d}{c}\ket{1}\right)
\eeql

(3) $a,d \neq 0$ or $b,c \neq 0$ , thus $b,c$ or $a,d$ must be non-zero. Product states can be taken the same form as above.

(4) $b,d \neq 0$  
\beql
\ket{\psi_{A_1}}&=&\frac{|b|}{\sqrt{|b|^2+|d|^2}}\left(\ket{0}+\frac{d}{b}\ket{1}\right)\\
\ket{\psi_{A_2}}&=&\frac{1}{\sqrt{|a|^2+|b|^2}}(a\ket{0}+b\ket{1})
\eeql

(5) $c,d \neq 0$, it is the same as the case (2).

It is simple to verify that for all of the cases
\beq
\ket{\psi_{A_1\!A_2}}=\ket{\psi_{A_1}}\otimes\ket{\psi_{A_2}}
\eeq
by means of  two conditions. One is separable condition $ad=bc$, another is the normalized condition $aa^*+bb^*+cc^*+dd^*=1$. So, theorem one is a sufficient criterion.  

Furthermore, we can use mathematical induction to prove the case of $n$-partite systems. Suppose this separability criterion is sufficient and necessary for $n-1$ partite systems. In the case of $n$-partite systems, by tracing off any single partite, a separable pure state in a $n-1$ partite system is obtained. Thus, the norm square of polarized vector of each partite for this state is always 1. Since the partite traced off is arbitrary, the criterion for $n$-partite systems must be valid. Theorem one is then necessary. Have assumed that for any $n-1$ partite the forms of product states can be written out. If we can not obtain the forms of product states for $n$ parties system which obey our criterion, it must be conflict with our precondition. Therefore, our criterion is sufficient. As an example , let us verify the case of a three partite system. Denoting
\beqa
\label{psi3q}
\ket{\psi_{A_1\!A_2\!A_3}}\!\!
&=&a\ket{000}+b\ket{001}+c\ket{010}+d\ket{011}\nonumber\\
& &+e\ket{100}+f\ket{101}+g\ket{110}+h\ket{111}
\eeqa
we can evaluate out
\beql
\!\bm{\xi}_{A_1}^2\!&=&\!1-4|a f - b e|^2-  4|a g - c e|^2-  4|a h - d e|^2\nonumber\\
& &-  4|b g - c f|^2-   4|b h - d f|^2-  4|c h - d g|^2\\
\!\bm{\xi}_{A_2}^2\!&=&\!1-4|a d - b c|^2-  4|a g - c e|^2-  4|a h - c f|^2\nonumber\\
& &-  4|b g - d e|^2-  4|b h - d f|^2-   4|e h - f g|^2\\
\!\bm{\xi}_{A_3}^2\!&=&\!1-4|a d - b c|^2-  4|a f - b e|^2-  4|a h - b g|^2\nonumber\\
& &-  4|c f - d e|^2-  4|c h - d g|^2-  4|e h - f g|^2
\eeql
Thus, the obviously independent expressions of our separability criterion in terms of coefficients of component of pure state are
\beqa
ad &=& bc, \quad af = be, \quad ah = bg, \quad ag =ce, \quad bg = cf,\nonumber\\
bh &=& df, \quad
cf = de, \quad ch = dg, \quad eh = fg
\eeqa
If $\ket{\psi_{A_1\!A_2\!A_3}}$ is separable, without loss of generality, in terms of our lemma we have
\begin{equation}
\ket{\psi_{A_1\!A_2\!A_3}}=(a_1\ket{0}+b_1\ket{1})\otimes(a_2\ket{0}+b_2\ket{1})\otimes (a_3\ket{0}+b_3\ket{1})
\end{equation}
Comparing with eq.(\ref{psi3q}), it follows that
\beqa
a=a_1a_2a_3,\ b=a_1a_2b_3,\ c=a_1b_2a_3,\ d=a_1 b_2 b_3\nonumber\\
e=b_1a_2a_3,\ f=b_1a_2b_3,\ g=b_1b_2a_3,\ h=b_2b_2b_3
\eeqa
Therefore, we have $\bm{\xi}_{A_1}^2=\bm{\xi}_{A_2}^2=\bm{\xi}_{A_3}^2=1
$. When $a,b\neq 0$, we obtain
\beql
\ket{\psi_{A_1\!A_2}}&=&\frac{|a|}{\sqrt{|a|^2+|c|^2+|e|^2+|g|^2}}\nonumber\\
& &\left(\ket{00}+\frac{c}{a}\ket{01}+\frac{e}{a}\ket{10}+\frac{g}{a}\ket{11}\right)\\
\ket{\psi_{A_3}}&=&\frac{1}{\sqrt{|a|^2+|b|^2}}(a\ket{0}+b\ket{1})
\eeql
Again, when the coefficients $c,e,g$ are all zero, 
\beq
\ket{\psi_{A_1}}=\ket{0}\quad \ket{\psi_{A_2}}=\ket{0}
\eeq
and if at least one element among the set $\{c,e,g\}$ is not zero, thus we can write down the product states  like two partite case. For example $c\neq 0$, we have
\beql
\ket{\psi_{A_1}}=\frac{|c|}{\sqrt{|c|^2+|g|^2}}\left(\ket{0}+\frac{g}{c}\ket{1}\right)\\
\ket{\psi_{A_2}}=\frac{|a|}{\sqrt{|a|^2+|c|^2}}\left(\ket{0}+\frac{c}{a}\ket{1}\right)
\eeql 
For the other cases, we can write down the product states in a similar way. In particular, if only one partite is separable with the other two partite, then the corresponding separable partite has a normalized polarized vector. For example, $A_3$-partite is separable with $A_1A_2$ partites, then $\bm{\xi}_{A_3}^2=1$. It must be emphasized that $A_2$ partite is separable with $A_1A_3$ partites means 
\beq
(I_{A_1}\otimes S_{A_2\!A_3})\rho_{A_1\!A_2\!A_3}(I_{A_1}\otimes S_{A_2\!A_3}^\dagger)=\rho_{A_1A_3}\otimes\rho_{A_2}
\eeq
where $S$ is an exchanging operator for the near partites,
\beq
S=\left(\begin{array}{cccc}
1&0&0&0\\
0&0&1&0\\
0&1&0&0\\
0&0&0&1
\end{array}\right)
\eeq
Therefore, we think that the partially separability, which means that  some partites are separable with the other partites, has the following criterion:

{\bf Conjecture One} The sufficient and necessary condition for partially separability of an arbitrary pure state in multipartite systems are that the polarized vectors of separable parties obey
\beq
\bm{\xi}_i^2=1,\quad\mbox{ $A_i$ takes over all of separable partites}
\eeq

We have checked it up to four partite systems, but the partially separable problem is still open to prove it for arbitrary partite systems in a strict method. 

{\bf Theorem Two}\  The sufficient and necessary conditions  for separability of an arbitrary pure state in $n$-partite systems made up of $n$ qutrits (every partite has a qutrit) are 
\beq
\label{scformp}
\bm{\xi}_{A_1}^2=\bm{\xi}_{A_2}^2=\cdots=\bm{\xi}_{A_n}^2 =\frac{4}{3}
\eeq
where $\bm{\xi}_{A_i}$ is a coherence vector of the reduced density matrix $\rho_{A_i}=\Tr_{\prod_{j=1,j\neq i}^n A_j}(\rho_{A_1\!A_2\cdots A_n})$ (tracing off $n-1$ partite), that is, its component is 
\beq
{\xi}_{A_i}^m=\left(\frac{\sqrt{6}}{2}\right)^{n-1}\!\!\!a_{0_10_2\cdots0_{i-1}m0_{i+1}\cdots 0_n }\ (m=1,2,\cdots,8)
\eeq
where subindex $0_j$ indicates 0 in the position $j$ and we have used the expansion of the density matrix of a qutrit
\beq
\rho_{A_i}=\frac{1}{3}I_{3\times 3}+\frac{1}{2}\sum_{m=1}^8{\xi}_{A_i}^m{\lambda}_m
\eeq

{\bf Proof}\  Let us start from bipartite systems. If denoting the pure state in a bipartite system as
\beqa
\label{psi2qt}
\ket{\psi_{A_1\!A_2}}&=&x_1\ket{11}+x_2\ket{12}+x_3\ket{13}+x_4\ket{21} +x_5\ket{22}\nonumber\\
& &+x_6\ket{23}+x_7\ket{31}+x_8\ket{32}+x_9\ket{33}
\eeqa
we can obtain 
\beqa
\label{2qtpvn2}
\bm{\xi}_{A_1}^2&=&\bm{\xi}_{A_2}^2=4/3-4|x_1 x_5 - x_2 x_4|^2\nonumber \\
& &- 4|x_1 x_6 -  x_3 x_4|^2- 4|x_1 x_8 -  x_2 x_7|^2 \nonumber\\
& & -4|x_1 x_9 -  x_3 x_7|^2-  4|x_2 x_6 -  x_3 x_5|^2\nonumber\\
& &- 4|x_2 x_9 -  x_3 x_8|^2-  4|x_4 x_8 -  x_5 x_7|^2\nonumber\\
& &- 4|x_4 x_9-  x_6 x_7|^2 -  4|x_5 x_9 -  x_6 x_8|^2
\eeqa
Thus, the sufficient and necessary conditions  for separability of an arbitrary pure state in bipartite systems of two qutrits can be 
written in form
\beqa
0&= &x_1 x_5 - x_2 x_4= x_1 x_6 - 
        x_3 x_4 =x_1 x_8 -  x_2 x_7\nonumber\\
&= & x_1 x_9 -  x_3 x_7=x_2 x_6 -  x_3 x_5=  x_2 x_9 -  x_3 x_8
\nonumber\\
& =&  x_4 x_8 -  x_5 x_7= 
  x_4 x_9-  x_6 x_7 =x_5 x_9 -  x_6 x_8
\label{scfor2qt}
\eeqa

It is easy to prove theorem two to be necessary. In fact, if the pure state $\ket{\psi_{A_1\!A_2}}$ is separable, in terms of our lemma, without of generality we have 
\begin{equation}
\ket{\psi_{A_1\!A_2}}=(a_1\ket{1}+b_1\ket{2}+c_1\ket{3})\otimes(a_2\ket{1}+b_2\ket{2}+c_2\ket{3})
\end{equation}
Comparing it with the equation (\ref{psi2qt}), up to an undetermined overall phase factor, we can obtain 
\beqa
\label{2qtsf}
x_1=a_1a_2,\quad x_2=a_1b_2,\quad x_3=a_1c_2,\\
x_4=b_1a_2,\quad x_5=b_1b_2,\quad x_6=b_1c_2,\\
x_7=c_1a_2,\quad x_8=c_1b_2,\quad x_9=c_1c_2,
\eeqa
Substituting eq.(\ref{2qtsf}) into eq.(\ref{2qtpvn2}), it immediately following that  our separable criterion is satisfied. 

It is a trivial case if only if there is one non-zero element in the coefficient set $\{x_i,i=1,2,\cdots,8\}$. Consider the case that at least two coefficients are not zero, for example $x_1,x_2\neq 0$, we have:  when $x_3,x_6,x_9=0$
\beqa
\ket{\psi_{A_1}}&=&\frac{|x_1|}{\sqrt{|x_1|^2+|x_4|^2+|x_7|^2}}\nonumber\\ & &\times\left(\ket{1}+\frac{x_4}{x_1}\ket{2}+\frac{x_7}{x_1}\ket{3}\right)\\
\ket{\psi_{A_2}}&=&\frac{1}{\sqrt{|x_1|^2+|x_2|^2}}\left(x_1\ket{1}+x_2\ket{2}\right)
\eeqa
when at least one element in the set $x_3,x_6,x_9$ is not zero, for example $x_3\neq 0$, $\ket{\psi_{A_1}}$ can be taken  the same form as above, but 
\beq
\ket{\psi_{A_2}}=\frac{1}{\sqrt{|x_1|^2+|x_2|^2+|x_3|^2}}\left(x_1\ket{1}+x_2\ket{2}+x_3\ket{3}\right)
\eeq
It must be emphasized that since separability conditions give out the relation among the coefficients, the expressions of product states can be written as the different forms.  Likewise, we can prove the other cases. In order to save space, we omit them. So, theorem two is a suffcient criterion.  As to the proof for $n$-partite systems, it is similar to one of theorem one. We do not intend to repeat it.  We also have checked the three partite systems made up of three qutrits. Except for the computation detail, there is no any new idea comparing with the three partite systems made up of three qubits. 

In general, we have

{\bf Conjecture Two}  In $n$-partite systems and the $i$th-partite with $r_i$ dimensional,  the sufficient and necessary conditions for separability of an arbitrary pure state are 
\beq
\label{scgf}
\bm{\xi}_{A_i}^2=2\left(1-\frac{1}{r_i}\right), \quad (i=1,2,\cdots,n)
\eeq
where $\bm{\xi}_{A_i}$ is a coherence vector of the reduced density matrix $\rho_{A_i}=\Tr_{\prod_{j=1,j\neq i}^n A_j}(\rho_{A_1\!A_2\cdots A_n})$ (tracing off $n-1$ partite) defined by
\beq
\rho_{A_i}=\frac{1}{r_i}I_{r_i\times r_i}+\frac{1}{2}\sum_{m_i=1}^{r_i^2-1}\bm{\xi}_{A_i}^{m_i}{\lambda}_{m_i}
\eeq
 $I_{r_i\times r_i}$ is $r_i$-dimensional identity matrix and $\lambda_{A_i}^{m_i}\ (m_i=1,2,\cdots, r_i^2-1)$ are generators of $SU(r_i)$. 

By using of our above idea and method, we can discuss the multipartite systems made up of qudits as well as the multipartite systems made up of mixture of qubits, qutrits and qudits. We can find our separability criterion to be valid. For example, $A_1$ partite is a qubit and $A_2$ partite is a qutrit. If denoting 
\beq
\label{psi2qt}
\ket{\psi_{A_1\!A_2}}=x_1\ket{01}+x_2\ket{02}+x_3\ket{03}+x_4\ket{11}+x_5\ket{12}+x_6\ket{13}
\eeq
It is easy to obtain
\beqa
\!\!1-\bm{\xi}_{A_1}^2\!& =&\!\frac{4}{3}-\bm{\xi}_{A_2}^2=4|x_1x_5-x_2x_4|^2\nonumber\\
& &+ 4|x_1x_6-x_3x_4|^2+4|x_2x_6-x_3x_5|^2
\eeqa
Of course, in the same way we can prove the above conjecture. However, how to prove our conjecture two strictly and generally is still open. 

In the end, we would like to point out that our result is helpful for defining measures of entanglement in multipartite and high dimensional systems. Actually, we can verify that for any cat states in multipartite systems,  the norm of coherent vector of  reduced density matrix of any $A_i$-partite takes the minimum value. Moreover, we have proved  that for a (fully) separable pure state, the norm of coherent vector of  reduced density matrix of any $A_i$-partite arrives at the maximum value. Therefore,  the entropy functions or their combinations or the other monotonic continuous functions varying with the norms of coherent vectors of  reduced density matrices of all of partites or their functions can be defined as the measures of entanglement for pure state in multipartite and high dimensional systems. Just like done and doing by us (limit within a scalar description and including measure of strength of entanglement degree of all of concerned partites) \cite{My2,My3,My4} based on Bennett,  Wootters {\it et.al}'s as well as Vedral's ideas \cite{Bennett,Wootters,Vedral} . Perhaps, not only the norm but also components of coherent vectors in a tensor (including vector) description of entanglement are required. While, distillation and preparation of entanglement for multipartite systems will act a role to judge what form is more reasonable and applicable.

\end{document}